\def\eq{\begin{eqnarray}}

\documentstyle [12pt] {article}
\begin{document}
\begin{titlepage}

\title{\Large{The Large-$N$ Limit of the Two-Hermitian-matrix model by
the
hidden BRST method} }
\vspace{1.2cm}

\author{
{\sc J. Alfaro}\thanks{Permanent address: Fac. de Fisica, Universidad
Cat\'olica de Chile, Casilla 306, Santiago 22, Chile.}
\\
      Theory Division, CERN\\ CH-1211,Geneva 23
}
\maketitle

\begin{abstract}

This paper discusses the large N limit of the two-Hermitian-matrix model
in zero dimensions, using
the hidden BRST method. A system of integral equations previously found
is
solved, showing that it contained the exact solution of the model in
leading
order of large $N$. \end{abstract}
\vfill
\vspace{11cm}
\begin{flushleft}
CERN--TH-6531/92 \\
July 1992
\end{flushleft}

\end{titlepage}

\newpage
 \section{INTRODUCTION}
Recently
there has been a considerable amount of progress in understanding the
physics of two-dimensional
quantum gravity coupled to $c\leq 1$ matter. This was brought by the
discovery of the double
scaling limit in matrix models \cite{gross}: The perturbative expansion of
the matrix model
provides a discretization of the two-dimensional surface,
classified according to its genus, which in the matrix model is a
given order in $1/N$ ($N$ being the range of the matrix). By tuning
the couplings in the matrix model in such a way that the perturbative
series diverges, one can approach the continuum limit of the two-dimensional
surface. In this way
it is possible to get a sum over all topologies  in  the  first  quantized
string,  using  as
a   discretization   of  the  two-dimensional  world  sheet  a
dense  sum  of Feynman  graphs  of  an  associated  matrix  model.
In  certain cases,  this  sum  can  be  calculated  exactly,  thus
providing  us with a definition of the non-perturbative string.

Critical exponents have been computed using this method.
Furthermore it has been shown that a Virasoro algebra naturally
arises in these matrix models \cite{verlinde}. This algebra is a realization
of the
Schwinger-Dyson equations in the $U(N)$ invariant set of operators of
the matrix model. Using this algebra, the equivalence of zero-dimensional
matrix models to topological
$2d$ gravity has been proved \cite{witten}. Very recently, an application of
these
ideas to compute intersection indices on moduli spaces of Riemann surfaces
has been
made \cite{kontsevich}.

It is clear that further progress in understanding the large-$N$ limit will
provide new grounds to
explore the physics of two-dimensional quantum gravity and non-critical
strings.

 Three years
ago  we  proposed  an  alternative  approach  to  the large-$N$ limit
based upon a  hidden  BRST  symmetry  which  is present in any
quantum  field  theory \cite{dam}.  In  particular  we  were
able to prove a theorem that permits the solution of the model in
leading order of large-$N$ in a very simple way \cite{alfaro}. An
important application of these ideas was made to the two-Hermitian-matrix
model \cite {metha},
which  corresponds  in  the  string  picture  to  the spherical topology of
the colored string  with
two  colors \cite{retamal}.  In such a case we were able to find a set
of integral equations that determine completely all correlations of
the traces of powers  of the two  matrices,  in  contrast  with
previous  results where only the partition function was
found \cite{kazakov}. Very recently, a $W_3$ algebra has emerged from this
model \cite{narain}.

 In this paper we study the solutions of these
equations in  a systematic way. We show that the system of integral equations
can be reduced to a
set of algebraic equations depending on some arbitrary constants. The
arbitrary constants are fixed
by the requirement of having either one or more cuts of an associated
analytic function.

Since our main objective is to test the hidden BRST method, we discuss in
detail
several simple examples, the simplest non-trivial case corresponding to a cubic
potential. We find that the solution of the problem is given by the same cubic
equation as in ref. \cite{narain}. To show explicitly how the arbitrary
constants are determined, we discuss the cubic potential with one free matrix.
As a final example we show how to solve the two-matrix model with a quartic
potential. In this case the system of integral equations reduces to a quartic
algebraic equation.

This paper is organized as follows: In section 2 we review the derivation of
the
Schwinger-Dyson equations using the hidden BRST method. As an illustration, in
section 3 we present a construction of the BRST charge encoding the Virasoro
constraints of the zero-dimensional, one-hermitian-matrix model. In section 4
we
derive the integral equations that describe the leading order in large-$N$
of the Two-Hermitian-matrix model. Section 5 show how to reduce
the integral equations  to an algebraic system. Section 7 contains the
examples,
and in section 8 we draw some conclusions.

\section{The Hidden BRST method}
\def\be{\begin{eqnarray}}
\def\ee{\end{eqnarray}}
In this section we review the BRST-invariant  derivation of Schwinger-Dyson
equations of
ref.\cite{dam}. Usually we study separately the invariances of the
classical action and the path integral
measure. In the hidden BRST method we prefer to write a BRST-invariant action
that reflects the symmetries of both the classical action and the functional
measure. To illustrate the point let us consider the following functional
integral describing the quantum properties of a bosonic field (A more general
case is considered in \cite{dam}):

\begin{eqnarray}
Z=\int[d\phi] exp[-S(\phi)] \label{n2}
\end{eqnarray}

Here $\phi(x)$ is a scalar field defined in the space-time point $x$.
The functional measure $[d\phi]$ is invariant under field translations:

\begin{eqnarray}
\phi(x)\rightarrow\phi(x)+\epsilon(x) \label{n1}
\end{eqnarray}

We want to build an action that is invariant under (\ref{n1}). In order
to do this we introduce a collective field $B(x)$ and write $Z$ (up to a
multiplicative constant) as follows:

\begin{eqnarray}
Z=\int[d\phi] [dB] exp[-S(\phi -B)]
\end{eqnarray}

$S(\phi -B)$ is invariant under (\ref{n1}) provided we transform $B$ at the
same
time by:

\begin{eqnarray}
B(x)\rightarrow B(x)+\epsilon(x)
\end{eqnarray}

The gauge invariant functions are $F[\phi-B]$ for any $F$.
In the gauge $B(x)=0$ we get back the original path integral (\ref{n2}).

In order to quantize the gauge invariant theory that we have just introduced we
use the BRST method\cite{mark}. The BRST and anti-BRST transformations are:
\begin{eqnarray}
\delta \phi    &= \psi  & \bar\delta\phi     = \bar\psi \nonumber\\
\delta B       &= \psi  & \bar\delta B       = \bar\psi \nonumber \\
\delta\psi     &= 0     & \bar\delta\psi     = -ib \nonumber \\
\delta\bar\psi &= ib    & \bar\delta\bar\psi = 0 \nonumber \\
\delta b       &= 0     & \bar\delta b       = 0 \label{n3}
\end{eqnarray}
To reach the gauge $B(x)=0$, we add the following term to the action:
\begin{eqnarray}
\delta\bar\delta[B^2] &=\delta[B\bar\psi] &=ibB-\bar\psi\psi
\end{eqnarray}

If we integrate over the fields $b$ and $B$ the action reduces to:
\begin{eqnarray}
S[\phi] +\int[dx] \bar\psi\psi\label{n4}
\end{eqnarray}

Using the equation of motion for $b$ and $B$ the BRST  and anti-BRST symmetry
(\ref{n3}) reduces to:
$$
\delta\psi(x) =\epsilon\psi(x)+\bar\epsilon\bar\psi(x)
$$
$$
\delta\psi(x) =\frac{\delta
S}{\delta\phi(x)}\bar\epsilon
$$
$$
\delta\bar\psi(x) =-\frac{\delta S}{\delta\phi(x)}\epsilon
$$
\begin{eqnarray}
\epsilon^2 =\bar\epsilon^2 &=\bar\epsilon\epsilon+\epsilon\bar\epsilon &=0
\label{n5}
\end{eqnarray}

We can easily check that(\ref{n5}) is indeed a symmetry of
(\ref{n4}). Associated to this symmetry is a set of Ward identities, for
instance, for any function $F$:
\begin{eqnarray}
<\delta[F(\phi)\psi(y)]> &=0
\end{eqnarray}
i.e.
\begin{eqnarray}
<\int dx \frac{\delta
F}{\delta\phi(x)}[\epsilon\psi(x)+\bar\epsilon\bar\psi(x)]\psi(y)+F\frac{\delta
S}{\delta\psi(y)}\bar\epsilon> &=0
\end{eqnarray}
Computing the (trivial) average with respect to the fermionic variables we get:
\begin{eqnarray}
<\frac{\delta F}{\delta\phi(x)}-F\frac{\delta S}{\delta\phi(x)}> &=0
\end{eqnarray}
The last identity encodes the most general Schwinger-Dyson equation of the
model.

Notice that the symmetry (\ref{n5}) conmutes with any symmetry of the classical
action of the model. In particular, if we study a theory which is invariant
under a large $N$ group, the invariance (\ref{n5}) will be respected order by
order in the $1/N$ expansion. We will use this important fact to derive a BRST
formulation of the Virasoro constraints satisfied by the zero dimensional
matrix model in the next section.

\section{Q CHARGE OF VIRASORO CONSTRAINTS}

By now it is well known \cite{verlinde} that the partition function of the
zero dimensional hermitian matrix model is annihilated by a set of operators
that satisfy a Virasoro algebra. The existence of this algebra is fundamental
to the proof of equivalence of two dimensional topological gravity and matrix
models\cite{witten} and to the connection between matrix models and
intersection indices in modular spaces of Riemann surfaces\cite{kontsevich}.

The Virasoro algebra we are discussing is a direct consequence of the
Schwinger-Dyson equations of the matrix model. So we expect that the
BRST-invariant derivation of these equations we presented in the previous
chapter will produce a BRST extension of the Virasoro algebra. This is indeed
the case as we are going to explain below. See also  (\cite{antal}
and\cite{jorge}):

Let us consider the generating function of $U(N)$ invariants correlations
functions of the zero-dimensional-hermitian matrix model:
\begin{equation}
Z[j]=\int[dM] exp[-\sum_{n=0}^{n=\infty}j_nTrM^n]
\end{equation}
Particular critical potentials are obtained by expanding around different
points in j-space.

 Using the procedure of the last section we form the
BRST-invariant extension of $Z[j]$:
\begin{equation}
Z[j,\eta,\bar\eta]=\int[dM] [d\psi][d\bar\psi] exp[-\sum_{n=0}^\infty(j_n TrM^n
+\eta_nTr\bar\psi M^n+\bar\eta_nTr\psi M^n)] exp[-Tr\bar\psi\psi]
\end{equation}
$\eta_n$ and $\bar\eta_n$ are anticonmuting sources.
The action $Tr\bar\psi\psi$ and the integration measure are invariant under a
nilpotent BRST transformation:
\begin{eqnarray}
\delta M=\psi & \delta\psi=0 & \delta\bar\psi=0
\end{eqnarray}
The Ward identity that $Z[j,\eta,\bar\eta]$ satisfies is:
\begin{equation}
<\sum_n j_n n Tr(M^{n-1}\psi)-\sum_n\eta_n\sum_m Tr(M^m\psi
M^{n-1-m}\bar\psi)>=0
\end{equation}
Replacing
\be
\psi \rightarrow -\frac{\partial exp[-Tr[\bar\psi\psi]
]}{\partial\bar\psi}
\ee
and integrating by parts over $\bar\psi$ gives:
\begin{equation}
Q Z[j,\eta,\bar\eta]=0
\end{equation}
with the BRST charge Q given by:
\begin{equation}
Q=\sum_{n=-1}^\infty \eta_n L_n-\sum_{n,m=-1}^\infty
\frac{n-m}{2}\eta_n\eta_m\frac{\delta}{\delta\eta_{n+m}}
\end {equation}
The Virasoro generators $L_n$ are given by:
\be
L_m=\sum_{n=0}^m \frac{\partial}{\partial j_{n+m}}\frac{\partial}{\partial
j_n}+
\sum_{n=0}^\infty n j_n\frac{\partial}{\partial j_{n+m}}
\ee
Notice that due to the identity $Tr 1=N$ we must replace
$$
\frac{\partial}{\partial j_0}\rightarrow -N
$$
These are the same $L_n$ found in \cite{luis}.
The nilpotency of $Q$ follows because the $L_n$ satisfy a Virasoro algebra:
\begin{equation}
[L_n,L_m]=(n-m)L_{n+m}  \ \ \ n,m=-1,0,1,\ldots
\end{equation}

The same procedure can be used to get a nilpotent BRST charge for the Virasoro
constraints of the large $N$ vector models discussed in \cite{poul}.
\vskip 1pc \noindent

 \section       {\bf THE INTEGRAL EQUATIONS}
\vskip 1pc

\def\be{\begin{equation}}
\def\ee{\end{equation}}
\def\beq{\begin{eqnarray}}
\def\eeq{\end{eqnarray}}

In this section we review the derivation of the system of integral equations
that describe the large-$N$ limit of the Two-Hermitian-matrix model in zero
dimension \cite{retamal}.

The action of the model is:
\be
S=\rm{Tr}\left[ V_1(M_1)+V_2(M_2)
    -cM_1M_2\right],
\ee
where $M_a$ are $N\times N$ Hermitian matrices and $V_i(M_i)$ are arbitrary
potentials.

We know that $S$ is invariant under $M_a\rightarrow UM_a U^\dagger
U\epsilon U(N)$.
Following section 2 \cite{dam} we form the BRST-invariant extension of $S$:
\be
\bar S=S+\rm{Tr}[\bar\psi_1\psi_1+\bar\psi_2\psi_2],
\ee
\noindent
where $\psi_a$ and $\bar\psi_a$ are Grassman valued $N\times N$ matrices;
$\bar S$ is
invariant under the following BRST transformation:
\beq
\delta M_a &=& \epsilon\psi_a+\bar\epsilon\bar\psi_a \\
\delta\psi_a &=& \bar\epsilon\delta S/\delta M_a \\
\delta\bar\psi_a &=& -\epsilon\delta S/\delta M_a \\
\epsilon^2
&=& \bar\epsilon^2=\epsilon\bar\epsilon+\bar\epsilon\epsilon=0.\label{n10}
\eeq
As explained in section 2, the Ward identities corresponding to this BRST
symmetry  are the Schwinger-Dyson equations of the model.
Moreover we can prove the following important proposition,
\cite{alfaro,retamal}:
\begin{quote}
\it{The $U(N)$ invariants of the quantum field are BRST invariants in leading
order of large-$N$.}
\end{quote}

The proof of this proposition relies heavily on the existence of the exact
symmetry of both the measure and the action provided by
(\ref{n10}).

In what follows, we are going to show how to use this proposition to solve the
Two-matrix model in leading order of large $N$.

 Let us write:
\be (M_a)_{  ij}=\sum_\alpha m_a^\alpha (T_a^\alpha)_{
ij}, \ee
where $m_a^\alpha$ are the eigenvalues of the matrix $M_a$ and $T_a^\alpha$ the
corresponding projectors satisfying:
\begin{eqnarray}
T_a^\alpha T_a^\beta=\delta_{\alpha\beta}\, T_a^\alpha &\sum_\alpha
T_a^\alpha=1\,
& Tr T_a^\alpha=1
\end{eqnarray}
\noindent
 then, since $m_a^\alpha$ are BRST invariants
,according to the proposition stated above, we get:
\be
\delta (M_a)_{ ij}^k=\sum_\alpha (m_a^\alpha)^k\delta (T_a^\alpha)_{ij}\,
\rm{with}\ \
\mbox{k=1,2...} \ee
\noindent
It follows that the BRST variation of the projectors is:
\be
\delta T_a^\alpha=\sum_{\gamma\neq\alpha}\frac{T_a^\alpha \delta M_a T_a^\gamma
+
       T_a^\gamma \delta M_a T_a^\alpha}{m_a^\alpha-m_a^\gamma}.
\ee

{}From the proposition, it follows that, in leading order of $1/N$,
 \be
\delta Tr[T^{\alpha_1}T^{\beta_1}\cdots T^{\alpha_n}T^{\beta_n}\psi_a]=0
\ee

After computing the fermionic average, we obtain:
\beq
(\Delta(x_{\alpha_1})-
cy_{\beta_n}-\sum_{\gamma\neq\alpha}\frac{1}{x_{\alpha_1}-x_\gamma})
T_{\alpha_1\beta_1\ldots\alpha_n\beta_n}=\nonumber\\
\sum_{k=2}^n\sum_{\gamma\neq\alpha_1}
T_{\alpha_1\beta_1\ldots\alpha_{k-1}\beta_{k-1}}
T_{\gamma\beta_k\alpha_{k+1}\beta_{k+1}\ldots\alpha_n\beta_n}
\frac{1-\delta_{\gamma\alpha_k}}{x_{\alpha_1}-x_\gamma}
+\sum_{\gamma\neq\alpha_1}
\frac{T_{\gamma\beta_1\alpha_2\beta_2\alpha_n\beta_n}}{x_{\alpha_1}-x_\gamma}
\eeq
We have introduced the definitions:
\beq
\Delta_1(x)=\frac{\partial V_1(x)}{\partial x}\\
\Delta_2(y)=\frac{\partial V_2(y)}{\partial y}\nonumber\\
T_{\alpha_1\beta_1\ldots\alpha_n\beta_n}=Tr T_1^{\alpha_1} T_2^{\beta_2}\ldots
T_1^{\alpha_n} T_2^{\beta_n}
\eeq
These
relations provide a whole set of identities that determine completely the
$U(N)$
invariants correlations of the model. Notice that the exact BRST symmetry of
the
model was very helpful in the derivation of the identities.
Although a complete set of identities follows from the previous equation, in
this paper we will discuss in detail the case $n=1$.

Taking the continuum version, $N\rightarrow\infty$, of the case $n=1$,we get:
\begin{eqnarray}
\Delta_1 (x)-2P\int  dx' u(x')/(x-x') &=& c\int dy v(y) y F(x,y) \label{0}\\
\Delta_2 (y)-2P\int  dy' v(y')/(y-y') &=& c\int dx u(x) x F(x,y)    \\
\lefteqn{P\int dx' u(x') F(x',y)/(x-x' )} \nonumber \\
 &=& [\Delta_1 (x) -cy-P\int dx' u(x')/(x-x' )]F(x,y) \label{1} \\
\lefteqn{P\int dy' v(y') F(x,y')/(y-y')} \nonumber \\
 &=&[\Delta_2 (y)-cx-P\int dy' v(y')/(y-y') ] F(x,y) \label{2} \\
\int dy' v(y') \Delta_2 (y')  F(x,y')  &=&cx         \\
\int dx' u(x') \Delta_1 (x') F(x',y) &=&cy           \\
\end{eqnarray}

\noindent
plus the normalization conditions:

\begin{eqnarray}
          \int  dx u(x)= \int dx u(x) F(x,y) = 1  \label{3}  \\
          \int  dy v(y) = \int dy v(y) F(x,y) = 1.    \label{4}
\end{eqnarray}
and $P$ stands for the principal value of the integral. $u$ ($v$) is the
density of eigenvalues of $M_1$($M_2$) and $F(x,y)$ is the continuum limit of
$Tr T_1^\alpha T_2^\beta$.
  Actually (\ref{1}) and (\ref{2}) plus the
normalization conditions (\ref{3}), and (\ref{4}) imply the other equations, as
can be easily seen by multiplying by $v(y)$ and integrating over $y$ in
(\ref{1}) to get (\ref{0}) .

\section{SOLUTION OF THE SYSTEM OF INTEGRAL EQUATIONS}

In this section we will show how to solve the integral system of last section.
Our  strategy will follow closely the standard procedure\cite{brezin}. We
will explain how the system of integral equations reduces to an algebraic
equation satisfied by an associated analytic function. The
different phases of the model are determined by the different number of cuts
the
analytic function may have.

Let us  introduce the following
functions:

\begin{eqnarray}
U(x,y) &=& \int_{d_2}^{d_1} dx' u(x') F(x',y)/(x-x' )     \\
H(x) &=&
\int_{d_2}^{d_1}  dx' u(x')/(x-x').
\end{eqnarray}
\noindent
These  functions  are  analytic  everywhere  in  the  complex  $x$
plane cut  along the real axis between $d_1$ and $d_2$. Across the
cut we have:

\begin{eqnarray}
(U_+ - U_-)/2 &=& -i\pi u(x) F(x,y)                \\
(U_+ + U_-)/2 &=& P\int dx' u(x') F(x',y)/(x-x' )    \\
(H_+ - H_-)/2 &=& -i\pi u(x)                            \\
(H_+ +H_-)/2 &=& P\int dx' u(x')/(x-x') \label{c1}
\end{eqnarray}

By the index $+$($-$) we indicate the value of the function in the upper(lower)
part of the cut.  Moreover, for large $x$ we have:

\begin{eqnarray}
U(x,y) \sim   1/x                                   \\
H(x)  \sim   1/x.     \label{a1}
\end{eqnarray}

          According to eq. (\ref{1}) we have:

\begin{equation}
U_-/U_+ = \frac{\Delta_1 (x) -cy -P\int dx' u(x')/(x-x') +i\pi u(x)}
{\Delta_1 (x) -cy - P \int dx' u(x')/(x-x') -i\pi u(x) } .
\end{equation}

 But $U(x,y)$ is analytic in the same region of the $x$ complex plane
than $H(x)$ is; therefore we must have:

\begin{equation}
U(x,y) = \frac{\lambda (x,y)}{[\Delta_1 (x) -cy -H(x)] }, \label{7}
\end{equation}
\noindent
with $\lambda (x,y)$ an integral function of $x$ which is a polynomial in
$x$ if $\Delta_1$  is a
polynomial also; $\lambda$ shall be chosen to get the right asymptotic
behavior for large
$x$. That is:

\begin{equation}
          U(x,y)\sim 1/x,        {\rm{for\,  large \,}} x
\end{equation}

Plugging (\ref{7}) in (\ref{c1}), we obtain

\begin{equation}
F(x,y) = \frac{\lambda (x,y)}{[(\Delta_1  - cy -H_+ )(\Delta_1  - cy -H_- )]},
  \label{5}
\end{equation}

In a similar way we can introduce analytic functions in the variable $y$:

\begin{eqnarray}
V(x,y) &=&\int_{e_2}^{e_1} dy' v(y') F(x,y')/(y-y') \\
I(y) &=&\int_{e_2}^{e_1} dy' v(y')/(y-y') ,
\end{eqnarray}
\noindent
These  functions  are  analytic  everywhere  in  the  complex  $y$
plane cut  along the real axis between $e_1$ and $e_2$. Across the
cut we have:

\begin{eqnarray}
(V_+ - V_-)/2 &=& -i\pi v(y) F(x,y)                \\
(V_+ + V_-)/2 &=& P\int dy' v(y') F(x,y')/(y-y' )    \\
(I_+ - I_-)/2 &=& -i\pi v(y)                            \\
(I_+ +I_-)/2 &=& P\int dy' v(y')/(y-y') \label{f1}
\end{eqnarray}

By the index $+$($-$) we indicate the value of the function in the upper(lower)
part of the cut.  Moreover, for large $y$ we have:

\begin{eqnarray}
V(x,y) \sim   1/y                                   \\
I(y)  \sim   1/y.     \label{e1}
\end{eqnarray}

 In  order to satisfy (\ref{2}) we must have:
\begin{equation}
V_-/V_+ = \frac{\Delta_2 (y) -cx- P\int dy' v(y')/(y-y')  + i\pi v(y)}
{\Delta_2 (y) -cx - P\int dy' v(y') /(y-y') -i\pi v(y)} .
\end{equation}
\noindent
Therefore we also have:
\begin{equation}
V(x,y)= \frac{\mu (x,y)}{[\Delta_2 (y)-cx- I(y)] }, \label{9}
\end{equation}
\noindent
with $\mu (x,y)$ being an integral function in $y$ chosen to match the
asymptotic
behavior of $V$ for large $y$.
We get
\begin{equation}
F(x,y)=\frac{\mu (x,y)}{[(\Delta_2 (y)-cx-I_+ )(\Delta_2  - cx - I_- )]}.
\label{6}
\end{equation}

Of course, (\ref{5}) and (\ref{6}) must coincide for $x,y$ in the support of
$u(x)$ and $v(y)$. This is a strong requirement,which permits us to find
$H$ and $I$, as will be clear after we work out some examples. The main
conclusion of this section is that the complicated integral system of the last
section has been reduced to a simpler equality between two functional forms for
$F(x,y)$. As far as I know this procedure to decouple an integral system was
not known before.

\vskip 1pc
\noindent
\section         {\bf EXAMPLES}
\vskip 1pc
In this section we want to discuss some simple models to illustrate
the method of section 5. We will see that for  given $\Delta_1$ and $\Delta_2$,
the consistency requirement for $F(x,y)$ will imply a set of equations for the
values on the cut of the analytic functions we have introduced in the last
chapter depending on some arbitrary constants. For polynomial $\Delta$'s,
this set of equations can be reduced to algebraic equations by using the Cauchy
theorem of residues. The arbitrary constants are the order parameters of the
different phases of the model. They are determined by the number of
disconnected cuts we allow in the analytic functions.

The simplest situation we want to discuss is the "free"
case. Namely:

\begin{eqnarray}
\Delta_1  &= x \\
\Delta_2  &= y .
\end{eqnarray}

Actually, we can compute $H$ and $I$ directly by replacing (15) and (16)
into (\ref{0}). By the usual method we find:

\begin{eqnarray}
u(x) &=& (1 -c^2)\sqrt {d^2 - x^2}/2\pi \\
d &=& 2/\sqrt {1-c^2} \\
v(y) &=& (1 -c^2 )\sqrt {d^2 - y^2}/2\pi .
\end{eqnarray}

Therefore:

\begin{eqnarray}
U(x,y)=\frac{1}{[x(1+c^2)/2 - cy + (1-c^2)\sqrt {x^2 -d^2}  /2\pi] }\\
V(x,y)=\frac{1}{[y(1+c^2)/2 - cx + (1-c^2)\sqrt {y^2 -d^2}  /2\pi]} .
\end{eqnarray}

The square root takes its principal value with a cut along the
real axis between $-d$ and $d$. It follows that:

\begin{equation}
F(x,y)=\frac{1}{[c^2 (x^2 + y^2 ) + 1 - c^2 - c(1 + c^2)xy]} . \label{8}
\end{equation}
\noindent
It  is  easy  to  check  that  (\ref{8})  gives  the  right  perturbative
expansion in c.

Let us examine a more general case:

\eq
\Delta_1(x) &=x+g_1x^2\\
\Delta_2(y) &=y+g_2y^2
\end{eqnarray}

According to (\ref{7}) and (\ref{9}) we must have:

\begin{equation}
U(x,y) = \frac{g_1x+\lambda_1(y)}{x+g_1x^2-cy-H(x)}
\end{equation}
\begin{equation}
V(x,y) = \frac{g_2y+\mu_1(x)}{y+g_2y^2-cx-I(y)}
\end{equation}

The comparison of the $F(x,y)$ coming from $U(x,y)$ and from $V(x,y)$ gives:

\eq
\mu_1c^2-cg_2(2\Delta_1- H_+-H_-) &=& a_2x^2+a_1x+a_0\\
g_2(\Delta_1-H_+)(\Delta_1-H_-) -c\mu_1(2\Delta_1- H_+-H_-) \nonumber\\
&=& b_2x^2+b_1x+b_0\\
\mu_1(\Delta_1-H_+)(\Delta_1-H_-) &=& g_1c^2x^3+h_2x^2+h_1x+h_0
\end{eqnarray}
\noindent
and

\eq
\lambda_1c^2-cg_1(2\Delta_2- I_+-I_-) &=& a_2y^2+b_2y+h_2\\
g_2(\Delta_2-I_+)(\Delta_2-I_-)-c\lambda_1(2\Delta_2- I_+-I_-)\nonumber\\
 &=& a_1y^2+b_1y+h_1\\
\lambda_1(\Delta_2-I_+)(\Delta_2-I_-) &=& g_2c^2y^3++a_0y^2+b_0y+h_0
\end{eqnarray}
\noindent
for certain constants $a_i,b_i,h_i$.

{}From equations (11) and (12), we get:

\begin{eqnarray}
\mu_1 & = & g_2[\Delta_1-(H_+-H_-)]/c+1 ,\\
\lambda_1 & = & g_1[\Delta_2-(I_+-I_-)]/c+1 ,
\end{eqnarray}
\noindent
from which we can fix several of the arbitrary constants:

\begin{eqnarray}
a_2 &=& -g_1g_2c \\
a_1 &=& -g_2c\\
a_0 &=& c^2\\
b_2 &=& -g_1c\\
h_2 &=& c^2
\end{eqnarray}
\noindent

The remaining system of equations is:

\begin{eqnarray}
H_+^2+H_+H_-+H_-^2-(H_++H_-)(2\Delta_1+c/g_2)+\nonumber\\
\Delta_1^2+2\Delta_1c/g_2+\bar
B &=& 0 \label{11}\\
(g_2(\Delta_1-(H_++H_-))/c+1)[\Delta_1-H_+][\Delta_1-H_-] &=&
g_1c^2x^3+c^2x^2+h_1x+h_0\label{12}
\end{eqnarray}
where we have defined
\begin {equation}
\bar B =(-g_1c^2x^2+b_1x+b_0)/g_2
\end{equation}
\noindent
To solve this system, multiply  eq.(\ref{11}) by $(H_+-H_-)$. A remarkable
property of the resulting expression is that it depends solely on the
discontinuities of powers of $H(x)$ along the cut. Using this property, we can
reduce the system (\ref{11}) and (\ref{12}) to a cubic equation for $H(x)$. In
fact, consider the integral:
$$
\frac{1}{2\pi i}\int dx \frac{H(x)^3}{x-y}
$$
along a curve that encloses the cut of $H(x)$. In the region of the complex
$x$ plane that does not contain the cut, the integral equals $H(y)^3$. On the
cut, we can use (\ref{11}) to express $H_+(x)^3-H_-(x)^3$ as a function of
$H_+(x)^2-H_-(x)^2$ and $H_+(x)-H_-(x)$. Then  the remaining integrals can be
computed in the region that do not contain the cut  to get the
following equation for $H$:

\begin{eqnarray}
{H(y)^3-(2\Delta_1+c/g_2)H(y)^2+F(y)H(y)-F'(y)-
F''(y)(<x>-y)/2-}\nonumber\\
F'''(y)(<x^2>-
2<x>y+y^2)/6- F^{(4)}(y)(<x^3>-3<x^2>y+\nonumber\\
3<x>y^2-y^3)/24+
{2g_1 =0}\label{13}
\end{eqnarray}
for certain additional constants $<x^i>$. We have introduced the notation:
\be
F(y) =\Delta_1^2+2c\Delta_1/g_2+\bar B
\ee
 The
condition (\ref{a1}) has been explicitly used to derive (\ref{13}), which
coincides with equation  (4) of ref. \cite{narain}.

Let us review what we have obtained up to here. The first step in the solution
of the set of integral equations of section 4 was to express  $U(x,y)$
($V(x,y)$) in terms of $H(x)$($I(y$) plus certain arbitrary functions of
$x$($y$). The next step was to impose the equality
\be
\frac{U_+(x,y)-U_-(x,y)}{H_+(x)-H_-(x)}=\frac{V_+(x,y)-V_-(x,y)}{I_+(x)-I_-(x)}
=
F(x,y)
\ee
 In this way, we found conditions that relate these arbitrary functions.
Remarkable enough, these conditions can be solved completely, reducing the
arbitrariness to a set of constants.This is
the main result of this paper. It
is clear that the same procedure will work for an arbitrary potential.

 In what follows, we will sketch that
solution of the  cubic equation which is relevant for the phase of the
Two-matrix
model containing the perturbative solution. That is, we will explore the
conditions under which $H(x)$ has only one cut on the real axis. This
requirement will fix all the arbitrary constants. In principle, we could find
all the phases of the model allowing more cuts, but we will not do this here.

 \subsection{Analysis of the cubic equation}

The cubic equation $H^3+a_2H^2+a_1H+a_0=0$ has the following solutions:

\begin{eqnarray}
H_n &=& w_n s_1+w_n^2s_2-a_2/3\\
s_1 &=& [r+(r^2+q^3)^{1/2}]^{1/3}\\
s_1s_2 &=& -q\\
q &=& a_1/3-(a_2/3)^2\\
r &=& (a_1a_2-3a_0)/6-(a_2/3)^3
\end{eqnarray}
\noindent
where $w_n$ is the n-th root of unity.

The equations (\ref{11}) and (\ref{12})  are satisfied if:
\begin{eqnarray}
q &=& \bar B/3+(\Delta_1-c/g_2)^2/9\\
r &=& c(g_1c^2x^3+c^2x^2+h_1x+h_0)/(2g_2)+q(\Delta_1-c/g_2)/2+\nonumber\\
(\Delta_1-c/g_2)^3/54\\
b_1 &=& g_1g_2-c(c^2+1)
\end{eqnarray}

The arbitrary constants will be determined by demanding that the analytic
continuation of the solution of the cubic that behaves as $1/x$ for large $x$
must have only two real branch points.

In our case this requirement is satisfied if $r^2+q^3$, which is a polynomial
in $x$ of degree 9, can
 be written as follows:
\be
r^2+q^3=-A(x^3+p_2x^2+p_1x+p_0)^2(x^3+s_2x^2+s_1x+s_0)\label{a4}
\ee
for some constants $p_i,s_i$ and $A$.

Actually  we can have different solutions of the problem by permitting a
larger number of branch
points \cite{Tan}, but we will not consider them here. In the next subsection
we study a simpler limiting case of the last equation.

\def\be{\begin{eqnarray}}
\def\ee{\end{eqnarray}}

\subsection{One of the matrices is free:$g_1=0$}
In this section we consider the limit $g_1=0$. The equation for $I(y)$
reduces to
a quadratic equation, which can be
 solved easily. The one cut solution implies the following values for the
arbitrary constants:
\begin{eqnarray}
h_1=0\\
bb_0=1
\end{eqnarray}
\noindent
and $h_0$ satisfies the following equation:

\begin{eqnarray}
64g_2^2h_0^3+h_0^2(96c^2g_2^2-96g_2^2-(c^2-1)^4)+\nonumber \\
h_0(30c^4g_2^2-60c^2g_2^2+30g_2^2-(c^2-1)^5)+ \nonumber \\
27g_2^4-c^6g_2^2+3c^4g_2^2-3c^2g_2^2+g_2^2 &=& 0 .\label{a3}
\end{eqnarray}

Now we want to check the consistency of the whole procedure by finding
$h_1,bb_0,h_0$ using (\ref{a4}). For $g_1=0$, $r^2+q^3$ is a polynomial of
degree 5; in order to get just two branch points, we must thus have:

$$
r^2+q^3=-A(x+p_0)^2(x^3+s_2x^2+s_1x+s_0) .
$$

We find the following equations to determine $p_0$ and $h_0$:
\begin{eqnarray}
h_0 = -({c^3} {\it g_2} + {c^5} {\it g_2} - 5 c {{{\it g_2}}^3} -
       6 {c^3} {{{\it g_2}}^3} + {c^4} {\it p_0} - 2 {c^6} {\it p_0} +
       {c^8} {\it p_0} - \nonumber \\
8 {c^2} {{{\it g_2}}^2} {\it p_0} -
       2 {c^4} {{{\it g_2}}^2} {\it p_0} - 12 {c^6} {{{\it g_2}}^2} {\it p_0} +
       {{{\it g_2}}^4} {\it p_0} -3 {c^3} {\it g_2} {{{\it p_0}}^2} -\nonumber
\\
       3 {c^5} {\it g_2} {{{\it p_0}}^2} + 12 {c^7} {\it g_2} {{{\it p_0}}^2} -
       6 {c^9} {\it g_2} {{{\it p_0}}^2} + 3 c {{{\it g_2}}^3} {{{\it p_0}}^2}
+
       30 {c^3} {{{\it g_2}}^3} {{{\it p_0}}^2} + \nonumber \\
       2 {c^2} {{{\it g_2}}^2} {{{\it p_0}}^3} +
       16 {c^4} {{{\it g_2}}^2} {{{\it p_0}}^3} -
       16 {c^6} {{{\it g_2}}^2} {{{\it p_0}}^3} -
       10 {c^3} {{{\it g_2}}^3} {{{\it p_0}}^4})/\nonumber  \\
     (3 {c^3} {\it g_2} + 9 {c^5} {\it g_2} + 9 c {{{\it g_2}}^3} +
       6 {c^2} {{{\it g_2}}^2} {\it p_0} - 36 {c^4} {{{\it g_2}}^2} {\it p_0} -
       6 c {{{\it g_2}}^3} {\it p_0}^2) ,
\end{eqnarray}
\begin{eqnarray}
0 = {c^2} {\it g_2} + 4 {c^4} {\it g_2} + 3 {c^6} {\it g_2} + {{{\it g_2}}^3}
+ \left( {c^3} + {c^5} - 5 {c^7} + 3 {c^9} - 12 {c^3} {{{\it g_2}}^2} \right)
     {\it p_0} +\nonumber \\
 \left( - {c^2} {\it g_2}  - 10 {c^4} {\it g_2} +
       11 {c^6} {\it g_2} \right)  {{{\it p_0}}^2} +
    8 {c^3} {{{\it g_2}}^2} {{{\it p_0}}^3}.
\end{eqnarray}
The last two equations for $p_0$ have a common root if $h_0$ satisfies
(\ref{a3}).

As usual the asymptotic expansion of the functions $H(x)$ and $I(y)$ around
$\infty$ gives  Green's functions of the eigenvalues of the corresponding
matrices. For instance, we get:
\be
<x> &=& \frac{c \left( -1 + {c^2} + { h_0} \right) }{
g_2}\nonumber\\
<x^2> &=& \frac{{c^2} - 2 {c^4} + {c^6} + {{{ g_2}}^2}
- {c^2} { h_0} +
        {c^4} { h_0}}{ g_2^2}
\ee

When $g_2\neq 0$, the system of equations to determine the constants in
(\ref{a4}) turns out to be too complicated to be solved analytically, so we
will not discuss it here.

Instead we will mention how these results generalize to a quartic potential.

\subsection{Quartic Potential}

For the sake of completeness, we will sketch the solution of the two-matrix
model with a quartic potential, which is defined by:
\be
\Delta_1=ax+g_1x^3\\
\Delta_2=by+g_2y^3,
\ee

So, we must have:
\be
U(x,y) &=& \frac{g_1x^2+\lambda_1 x+\lambda_0}{g_1x^3+ax-cy-H(x)}\\
V(x,y) &=& \frac{g_2y^2+\mu_1y+\mu_0}{g_2y^3+by-cx-I(y)}
\ee

Again, imposing the equality of $F(x,y)$ coming from the two previous
functions,
we get a system of algebraic equations to determine the  arbitrary functions
$\lambda_0,\lambda_1,\mu_0,\mu_1$.One of these equations is:
\be
g_2(H_+^4-H_-^4)-(3g_2\Delta_1+c)(H_+^3-H_-^3)+
(3g_2\Delta_1^2+B+3c\Delta_1)(H_+^2-H_-^2)-
\nonumber\\\Delta_1(g_2\Delta_1^2+2B+3c\Delta_1)(H_+-H_-)=Dc(H_+-H_-)
\ee
\noindent
with $$
B=b_2x^2+b_1x+b_0$$ and $$
D=-g_1cbx^3+d_2x^2+d_1x+d_0
$$

 This system can be solved using  the Cauchy
theorem of residues as was used to derive (\ref{13}). We get a quartic
equation for $H(x)$ and a similar one  for $I(y)$.

Again the complicated set of integral equations has been reduced to an
algebraic equation, depending on a set of arbitrary constants. To completely
determine the arbitrary constants we may demand that $H(x)$ has only one cut
in the $x$ complex plane. Once this choice has been made all
$U(N)$-invariant correlations of the two matrices which involve $u(x)$,$v(y)$
and $F(x,y)$ can be computed.

 \section{CONCLUSIONS AND OPEN PROBLEMS}

We have been able to solve the system of integral equations that describe the
large-$N$ limit of the two-matrix model( a particular subset of the
Schwinger-Dyson equations) by reducing it to an algebraic equation
satisfied by an associated analytic function which depends on some arbitrary
constants.  In
particular, we have verified that the solution found in \cite{narain} is
obtained. In certain simple cases, we have been able to find these constants
explicitly by choosing the analytic function with only one cut (the
perturbative phase of the model). In more general situations, finding the
constants for the one cut solution analytically is very complicated.

 Our solution is appropriate to explore all the different phases of the
model which are characterized by the set of arbitrary constants.
 The
constants are no longer arbitrary if we fix the number of branch points of the
associated analytic function. We see this result as  a strong test of the
hidden
BRST method.

It is an interesting open problem to find the other solutions of (32),
which will enable us to compute more general correlations functions of the
form $Tr M_1^{n_1} M_2^{m_1} M_1^{n_2}M_2^{m_2}\ldots$.

\section*{ACKNOWLEDGEMENTS}

The author wants to thank Luis Alvarez-Gaum\'e  for useful
discussions and Spenta Wadia for having called  his attention to ref.
\cite{narain}. He also wants to express his gratitude to P.H. Damgaard and
G. Shore for a critical reading of the manuscript.
His work has been supported by
a EC CERN Fellowship. He also acknowledges help from the Fundaci\'on Andes \#
C-11666/4.

 \end {document}